\documentclass[%
superscriptaddress,
twocolumn,
 amsmath,amssymb,
 aps,
]{revtex4-2}

\usepackage{graphicx}
\usepackage{dcolumn}
\usepackage{bm}
\usepackage{amsmath,amssymb,amsthm,bm,mathtools,amsfonts,mathrsfs,bbm,dsfont} 
\usepackage{graphicx}
\usepackage{braket}
\usepackage{color}
\usepackage{float}
\usepackage[skins,theorems]{tcolorbox}
\usepackage{soul}
\usepackage[paperwidth=210mm,paperheight=297mm,centering,hmargin=1.5cm,vmargin=3.5cm]{geometry}

\usepackage[unicode=true, breaklinks=false, pdfborder={0 0 1}, backref=false, colorlinks=true, linkcolor=blue, citecolor=blue, urlcolor=blue]{hyperref}

\renewcommand{\vec}[1]{\boldsymbol{#1}}
\newcommand{\vB}{\vec{B}}

\newcommand{\vn}{\vec{n}}

\newcommand{\ve}{\mathbf{e}}

\newcommand{\la}{\langle}
\newcommand{\ra}{\rangle}

\newcommand{\da}{\dagger}

\newcommand{\Op}[1]{\hat{#1}}

\newcommand{\osigma}{\Op{\sigma}}

\newcommand{\oH}{\Op{H}}

\newcommand{\oU}{\Op{U}}

\newcommand{\vmu}{\vec{\mu}}
\newcommand{\id}{\ensuremath{\mathbbm 1}}

\newcommand{\ovsigma}{\Op{\vec{\sigma}}}
\newcommand{\ovmu}{\Op{\vec{\mu}}}

\begin{document}

\preprint{APS/123-QED}

\title{Spin Resonance Spectroscopy with an Electron Microscope}

\author{Philipp Haslinger}
\email{philipp.haslinger@tuwien.ac.at}
\affiliation{Vienna Center for Quantum Science and Technology, Atominstitut, TU Wien, Stadionallee 2, 1020 Vienna, Austria}
\affiliation{University Service Centre for Transmission Electron Microscopy,TU Wien, Wiedner Hauptstraße 8-10/E057-02, 1040 Wien, Austria}
\affiliation{Erwin Schr\"{o}dinger International Institute for Mathematics and Physics, University of Vienna, 1090 Vienna, Austria}

\author{Stefan Nimmrichter}
\email{stefan.nimmrichter@uni-siegen.de}
\affiliation{Naturwissenschaftlich-Technische Fakultät, Universität Siegen, Walter-Flex-Str.~3, 57068 Siegen, Germany}

\author{Dennis R\"atzel}
\email{dennis.raetzel@zarm.uni-bremen.de}
\affiliation{Institut f\"ur Physik, Humboldt-Universit\"at zu Berlin, Newtonstraße 15, 12489 Berlin, Germany}
\affiliation{ZARM, Unversität Bremen, Am Fallturm 2, 28359 Bremen, Germany}
\affiliation{Erwin Schr\"{o}dinger International Institute for Mathematics and Physics, University of Vienna, 1090 Vienna, Austria}


\begin{abstract}
Coherent spin resonance methods, such as nuclear magnetic resonance and electron spin resonance spectroscopy, have led to spectrally highly sensitive, non-invasive quantum imaging techniques. Here, we propose a pump-probe spin resonance spectroscopy approach, designed for electron microscopy, based on microwave pump fields and electron probes. We investigate how quantum spin systems couple to electron matter waves through their magnetic moments and how the resulting phase shifts can be utilized to gain information about the states and dynamics of these systems. Notably, state-of-the-art transmission electron microscopy provides the means to detect phase shifts almost as small as that due to a single electron spin. This could enable state-selective observation of spin dynamics on the nanoscale and indirect measurement of the environment of the examined spin systems, providing information, for example, on the atomic structure, local chemical composition and neighboring spins.
\end{abstract}

\maketitle

Modern-day transmission electron microscopy (TEM), with advanced techniques for aberration correction and cryogenic sample preparation \cite{Reimer2008Transmission,Bai2015}, is a well-matured technology that employs wave properties of electrons to resolve structures at an atomic level. The development of ultra-fast transmission electron microscopy ($< 1$ ps) made it possible to investigate processes with both nanometer spatial resolution and sub-picosecond temporal resolution \cite{Lobastov2005, Zewail2006, Vanacore2018}, nowadays in optimized interferometric setups \cite{Houdellier2019,arbouet2018ultrafast,Feist2017,Tauchert2022,Houdellier2018}. Most ultra-fast pump-probe experiments are based on a laser-triggered sample excitation (UV-IR) followed by a highly temporally resolved, but sparse electron probe.
At the same time, bright electron sources with beam blanking systems ($<10$ ns ) \cite{Zhang2020} and fast direct electron detectors ($\sim260$ ps) \cite{llopart2022timepix4} have been integrated into aberration-corrected TEMs. 
This facilitates the time-resolved probing of fast processes such as quantum spin dynamics.

The default method to probe the dynamics of spin samples is electron spin resonance (ESR) and nuclear magnetic resonance (NMR) spectroscopy. 
This non-invasive imaging technology \cite{Liang2000, Callaghan1993Principles} revolutionized not only medical diagnostics \cite{Bullmore2012}, biology \cite{Borbat2001, sivelli2022nmr}, and chemistry \cite{Weil1994, Wertz2012}, but also high-precision measurements for fundamental physics \cite{Ramsey1950}, including searches for dark energy and dark matter \cite{Safronova2018}. Magnetic resonance spectroscopy 
is also used to characterize electrode materials for electrochemical energy storage \cite{Nguyen2020,Zhao2021}, meeting the challenges of the green energy revolution.

In this paper, we propose a microwave-pump electron-probe method for ESR and NMR down to the level of individual spins, with coherence times on the order of 100 ns up to minutes or even hours \cite{Schweiger2001, Callaghan1993Principles}. 
While standard ESR (NMR) techniques employ magnetic field gradients and optimized antennas to overcome the diffraction limit imposed by the microwave probe fields \cite{Anders2018}, we point out the surprisingly low technical challenge towards probing spin dynamics directly at the nanoscale with highly controlled electron beams in a TEM.   

In the following, we present a basic electron-interferometric scheme that implements the proposed method. Therein, the magnetic moment of a spin sample imprints a phase onto the wave function of a passing electron. A tailored microwave pump pulse interaction causes a controlled change of that spin and thus a change in phase, which can then be detected by interferometric phase imaging \cite{Lichte2008,boureau2020high,Ishikawa2023}. The microwave pulse thereby enables a differential measurement technique for separating the phase contribution of the spins.
As an alternative to the interferometric read-out, one could measure the spatial gradient of the phase, which manifests in a deflection of the electron beam detectable in the far field, as in Differential Phase Contrast (DPC) imaging \cite{Ishikawa2023,boureau2020high,Edstrom2019}.

In magnetic resonance spectroscopy, one investigates the local resonance frequency of target spins by resonantly driving the transition of the two spin states with a tunable microwave pulse. The resonance can be controlled by an external magnetic bias field $\vB_0$, but its precise value crucially depends on the spins' immediate surroundings---which one can thus infer from a measurement of the magnetic moment after the excitation pulse. 

\begin{figure}
    \includegraphics[width=8cm,angle=0]{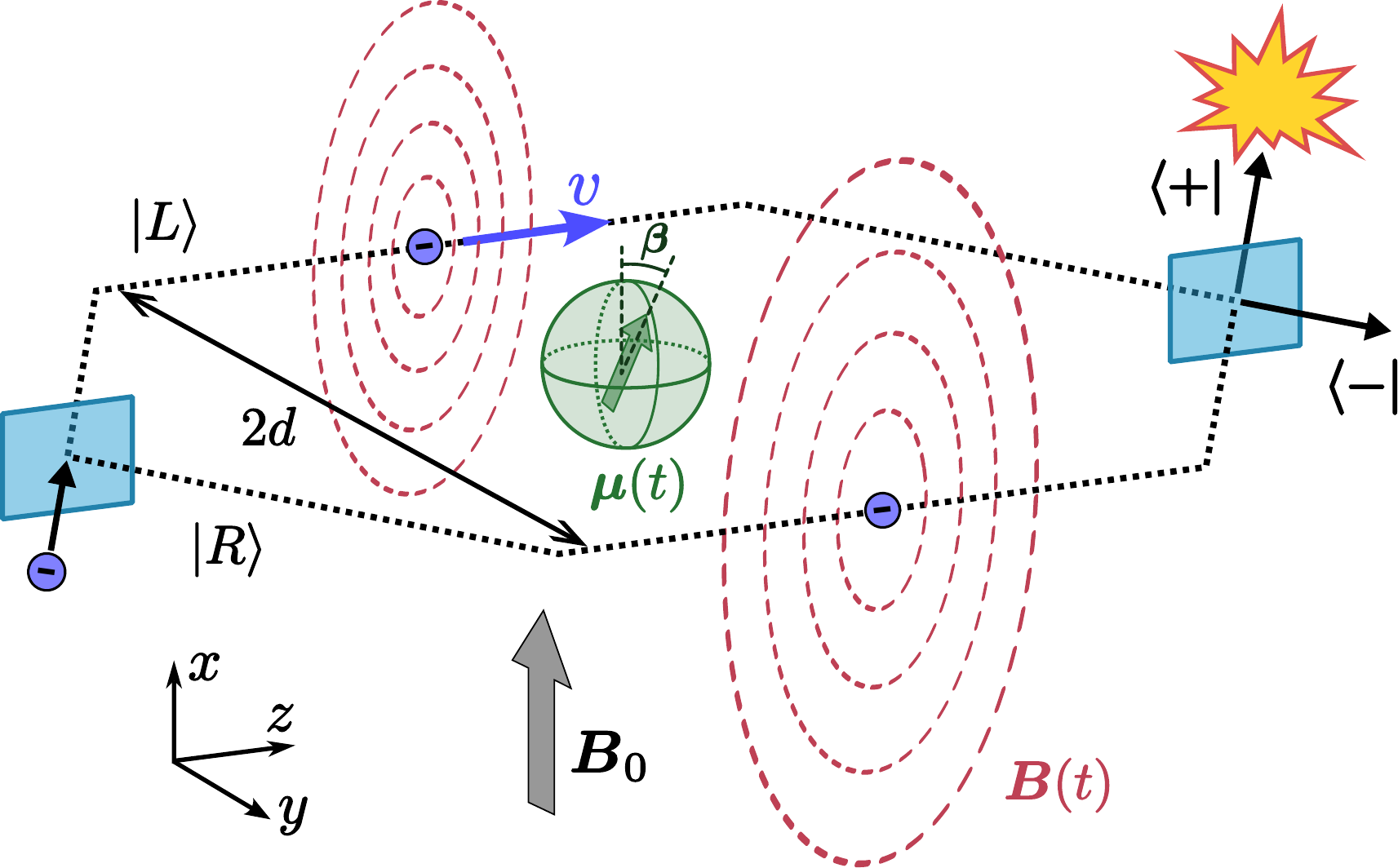}
    \caption{Model of the Electron Microscopic Spin Resonance Spectroscopy (EMSRS) scheme based on a Mach-Zehnder interferometer configuration. An electron beam is split coherently into two arms separated by the distance $2d$, such that each individual electron can pass to the left side (path state $|L\ra$) and to the right side ($|R\ra$) of a small spin sample in the center. The sample is aligned with respect to a bias field $\vB_0$, and is described by a magnetic moment $\vmu(t)$, which interacts with the path-dependent magnetic field $\vB(t)$ of the moving electron and thereby causes a small phase shift between the two arms. The phase shift is detected through the difference in electron counts upon recombination of the two arms at a second beam splitter. Analogous to ESR/NMR, microwave pulses may be applied to toggle between spin orientations of the sample and enable differential measurements to subtract systematic phase shifts.
    \label{fig:MZI}}
\end{figure}

Here, the change of the magnetic moment is monitored through free-space electrons. 
For their interferometric readout, consider the Mach-Zehnder interferometer (MZI) model configuration sketched in Fig.~\ref{fig:MZI}. A dilute beam of non-interacting electrons propagating with velocity $v$ in the $z$-direction is coherently split into two distinct arms separated by $\Delta y = 2d$, each passing by a magnetic sample on opposite sides. The sample is described by a magnetic moment $\vmu (t)$ placed between the two arms in the $yz$-plane, with a time-dependent orientation.
We assume that the electron beam  is split much further away from the sample than the closest distance at which the electrons pass it. The sample will interact with the moving electron and thereby build up a relative phase between the interferometer arms, which will change the likelihood that the electron is detected in one of the output ports upon recombination of the two arms. This is the measurement signal by which we seek to monitor changes in the sample's magnetic moment after it is manipulated by a microwave pulse.

\begin{figure*}
\centering
\includegraphics[width=14cm,angle=0]{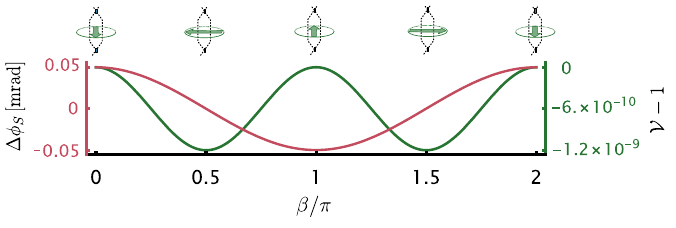}
\caption{\label{fig:visibility-phase} {Phase shift $\Delta\phi_S$ (red) and entanglement-induced drop in interferometric visibility $\mathcal{V}-1$ (green) due to a single electron spin in a pure state (on the surface of the Bloch sphere) at distance $d=0.1\,$nm in the Mach-Zehnder electron interferometer configuration in Fig.~\ref{fig:MZI}. The spin expectation value is oriented at varying angles $\beta$ with respect to the normal axis of the interferometer plane. When the spin lies in plane ($\beta=\pi/2,3\pi/2$), the phase shift vanishes and the visibility reaches its minimum. Orthogonal to the plane ($\beta=0,\pi$), the phase shift is maximal at full visibility.
Using microwave pulses to toggle between $\beta=0$ and $\pi$, one can observe the differential phase shift $2\Delta\phi_S\approx 0.11\,$mrad.}  } 
\end{figure*}

Treating the magnetic dipole moment as a classical source of a magnetic field, the differences of the phases imprinted on the electronic matter wave when passing on each side of the source is given to leading order by the magnetic flux through the interferometer plane \cite{Lichte2008,mccray2021understanding},
\begin{equation}\label{eq:deltaPhiS_A}
    \Delta \phi_S = -\frac{e}{\hbar}\Phi_\mathrm{mag} = -\frac{e}{\hbar}\oint d\vec{s} \cdot \vec{A} (\vec{s})\,.
\end{equation}
where $e$ is the electron charge and $\hbar$ is the reduced Planck constant.  $\Delta \phi_S $ is the well-known magnetic Aharonov-Bohm phase \cite{Aharonov1959,tonomura1982observation}; the integral over the vector potential $\vec{A}$ is performed on the closed curve given by the semiclassical electron trajectories corresponding to the two arms. In the case of a point-like magnetic dipole moment of magnitude $\mu$ that is oriented in the $\ve_x$-direction normal to the interferometer plane and placed exactly in the middle between the two arms of the MZI, we can approximate the vector potential by
\begin{eqnarray}\label{eq:vecpot}
	\vec{A} (\vec{r}) \approx \frac{\mu_0 \mu}{4\pi} \frac{\ve_x\times\vec{r}}{r^3}\, ,
\end{eqnarray}
where $r=\sqrt{x^2+y^2+z^2}$ and $\mu_0$ is the magnetic permeability. This results in the phase shift (see Appendix~\ref{sec:phase})
\begin{equation}\label{eq:DeltaphiS}
    \Delta \phi_S =  \frac{e\mu_0 \mu}{\pi\hbar d} \,.
\end{equation}
A single electron spin exhibits a magnetic dipole moment of approximately one Bohr magneton, $\mu = -g_s \mu_B/2 \approx -e\hbar/2m_e$.
In TEM however, one normally analyzes samples with a thickness of around 10-300 nm, corresponding to atomic columns of up to 1000 spin-active atoms with their associated magnetic moments. For classical spherical sources, we obtain an additional geometrical factor of $\pi/2$ (see Appendix~\ref{sec:magpolT}). 
Table \ref{tab:phaseShifts} compares the phase shift differences (between both interferometer-plane-orthogonal spin orientations) induced by a single electron spin, a single nuclear spin, and spin columns when the interfering electrons pass at distances of $0.1\,$nm and $1\,$nm. As we will discuss below, state-of-the-art electron-phase sensitive methods are approaching the necessary resolution to detect such tiny phase shifts caused by single quantum spins---rendering our proposed method feasible.

Such a level of sensitivity comes with an important side effect. If the magnetic moment is associated with a quantum system and if it is not orthogonal to the interferometer plane, the back-action on the spin system has to be included. An appropriate treatment can be given by reversing the perspective and describing the action of the electron's magnetic field on the quantum system. In this description, each electron is a moving charge that interacts with the magnetic moment represented by the dipole operator $\hat\vmu (t)$ through the magnetic field $\vB (t)$ it creates at the sample location. We focus here on the quantum regime of a single spin-$1/2$ aligned with respect to a bias field and driven by a short microwave
pulse in order to prepare a desired input state. To this end, we replace $\hat\vmu(t)$ by a spin operator, $    \ovmu(t) = \mu\, \ovsigma(t)$ that freely precesses at the Larmor frequency $\omega_0 = \mu B_0/\hbar$ about the magnetic field axis $\vn$.
Details can be found in Appendix \ref{sec:magfieldonsample}. We restrict our considerations to the stroboscopic regime where the passage of the electron through the interaction region with $\hat\vmu(t)$ is much shorter than the time scale of the free dynamics of the spin system (the inverse Larmor frequency or the life-time of the spin state). We may thus approximate $\hat\vmu(t) \approx \hat\vmu(t_0)$ over the duration of the single-electron pulse.

 We neglect the energy loss or gain of the electron in the interaction with the spin system as well as flips of the electron spin (see Appendix~\ref{sec:posDOFs}). The interaction is then approximately given by the magnetic field of the electron at the position of the spin system integrated over the semi-classical trajectory of the electron passing by the sample at distance $d$ in the MZI corresponding to the two arm states $|L\ra$ and $|R\ra$. Due to the orientation of the setup  (see Fig. \ref{fig:MZI}), the integrated magnetic field is oriented in the $x$-direction and its effect on the spin system is given by $x$-projection of the spin operator. 
We consider the initial superposition state $|\phi\ra = (|R\ra + e^{i\phi}|L\ra)/\sqrt{2}$ created by the first beam splitter of the MZI and an adjustable external phase shift $\phi$. The interaction with the spin system leads to additional phase shifts and may entangle the electron with the spin system depending on the initial state of the latter. 
Ultimately, the electron is detected at one of the two output ports of the second beam splitter $|+\ra =(|R\ra + |L\ra)/\sqrt{2}$ and $|-\ra = (|R\ra - |L\ra)/\sqrt{2}$  say, $|+\ra$.  The likelihood for this to happen,
\begin{equation}
    \begin{aligned}
        \label{eq:likelihood}
    p_+ (\theta,\phi; t_0) &= \frac{1}{2} \left[ 1 + \mathcal{V} \cos(\phi - \Delta \phi_S) \right],\, \mathrm{where} \\
    \Delta\phi_S &= \mathrm{arctan}\left(\la \osigma_x(t_0)\ra \tan2\theta\right) \,,
    \end{aligned}
\end{equation}
carries information about the sample's average spin-$x$ component at the probe time, $\la \osigma_x(t_0)\ra  = \la \psi|\osigma_x(t_0)|\psi\ra$. Here, we introduce the interaction strength $\theta = e \mu_0 \mu/2\pi \hbar d$, and we denote expectation values with respect to the sample state by $\la \cdot \ra$, noting that the results extend to classical mixtures of pure states. 
The effect of the sample on the sinusoidal interference signal described by $p_+ (\theta,\phi;t_0)$ is two-fold: Firstly, it shifts the sinusoidal fringes by the net phase $\Delta\phi_S$. Secondly, it reduces the fringe contrast, or visibility $\mathcal{V}$. For $\la \osigma_x(t_0)\ra=\pm 1$, we find $\Delta\phi_S = \pm 2\theta$ at $\mathcal{V}=100\,\%$ visibility. This resembles the ideal, backaction-free sensing of a fixed magnetic moment through the induced phase shift $2\theta$ recovering $\Delta\phi_S$ in equation \eqref{eq:DeltaphiS}.

In the general case, the $x$-component $\la \osigma_x(t_0)\ra$ of the spin vector oscillates with the Larmor precession frequency $\omega_0$ as a function of $t_0$ (unless the bias field $\vB_0$ points along $\vn = \ve_x$). Hence, by periodically probing the sample at a fixed value of $\omega_0 t_0$ modulo $2\pi$ and recording the electron counts with adjustable external phase $\phi$, one obtains sinusoidal interference fringes with the predicted phase shift $\Delta\phi_S$ and visibility $\mathcal{V}$.
In figure \ref{fig:visibility-phase}, we consider a pure state of the spin system and plot $\Delta\phi_S$ and $\mathcal{V}$ as functions of the angle $\beta=\arccos\la \osigma_x(t_0)\ra$ between the normal to the interferometer plane and the expectation value of the spin vector. There are two extreme cases: 1) $\beta = (n+1/2)\pi $ with $n$ integer, where the phase vanishes and the visibility of the interference pattern reaches its minimum. This case corresponds to the situation of the spin vector lying in the interferometer plane. 2) $\beta = n\pi $, where the phase is maximal and visibility is full. In this case, the spin vector is orthogonal to the interferometer plane. 

If we assume that we measure the current at the two output ports separately, we can infer the interaction strength $\theta$ from the mean current difference, which is proportional to $2p_+(\theta,\phi;t_0)-1$ from \eqref{eq:likelihood}. The measurement uncertainty, given in the limit of many repetitions by the variance of the current difference, translates accordingly to a mean-square error $\mathrm{Var}[\theta]$. We conclude that the maximal precision that can be reached by the intensity measurement at the two ports is $\sqrt{\mathrm{Var}[\theta]}=1/2\sqrt{N_e}$ for $N_e$ non-interacting electrons in the interferometer (see Appendix~\ref{sec:measprec}). 

\begin{figure}
   \includegraphics[width=8cm,angle=0]{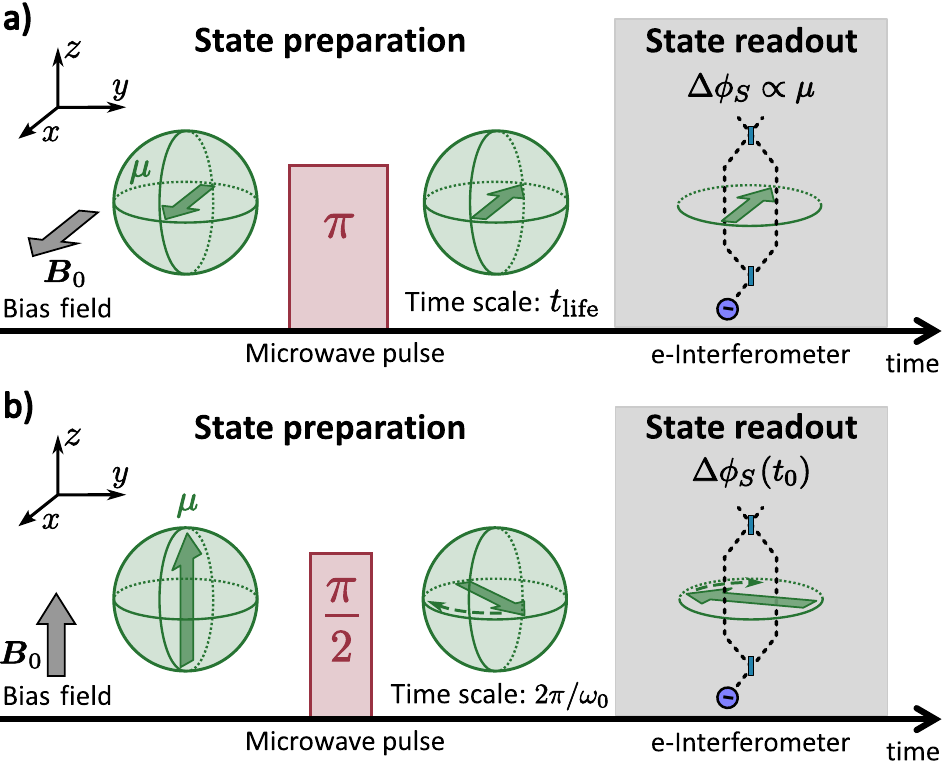}
    \caption{Electron interferometric schemes to investigate spin dynamics:
A quantum spin is aligned with respect to a bias field $\vB_0$ (on the Bloch sphere). A microwave pump
pulse manipulates the spin and causes a change of the associated magnetic moment. The resulting variation of magnetic flux in $x$-direction is then measured with an interferometric electron probe in the $yz$-plane. 
(a) When the bias field is perpendicular to the interferometer plane, a $\pi$-pulse toggles between the two spin eigenstates and thus allows to detect a constant phase shift during the relatively long lifetime $t_{\rm life}$ of the excited state.
(b) For a bias field aligned in the plane, a $\pi/2$-pulse causes the spin to precess about the field axis, allowing to probe a fast varying phase shift with a precession time $2\pi/\omega_0 \ll t_{\rm life}$. Only an ultra-fast TEM might be able to resolve the phase shift. For slow electron detection, this varying phase shift will on average reduce the interference visibility.\label{fig:preparation}} 
\end{figure}
With the electron readout scheme at hand, we now discuss two exemplary spectroscopy protocols on the spin sample, as depicted in Fig.~\ref{fig:preparation}. In (a), we let the bias field point in $\ve_x$-direction so that the interferometer probes the time-independent spin polarization $\la\osigma_x(t_0)\ra = \la \osigma_x \ra = s_x$. This causes an average phase shift by $\arctan(s_x \tan 2\theta )$, accompanied by a visibility drop if the spin state is mixed ($|s_x|<1$), for example, due to thermal excitation. After interaction with a microwave  $\pi$-pulse, the spin and therefore the phase shift are reversed, allowing for a differential phase measurement.

In Fig.~\ref{fig:preparation}(b), we consider a strong bias field $\vB_0$ that sets the spin quantization axis as $\vn=\ve_z$, such that $\osigma_x (t_0) = \osigma_x \cos \omega_0 t_0 + \osigma_y \sin \omega_0 t_0$. An initially $\ve_z$-aligned or anti-aligned spin state, or a mixture of these with $\la \osigma_z \ra = s_z \in [-1,1]$ and $\la \osigma_{x,y}\ra = 0$, will cause a visibility drop by $|\cos 2\theta|$ due to the entanglement between electron and spin state by the interaction, and there will be no net phase shift. If we then apply a $\pi/2$ microwave pulse, the spin will precess with $\la \osigma_x (t_0)\ra = s_z \sin \omega_0 t_0 $, resulting in an oscillating phase shift and visibility. In the optimal case of a pure spin state ($s_z = \pm 1$), the phase periodically shifts as much as $\pm 2\theta$ at full visibility. These maximal absolute phase shifts are obtained for $t_0$ being an integer multiple of $\pi/\omega_0$, in other words, when the precessing spin vector is orthogonal to the interferometer plane. From this result, we can also conclude that the integrated signal from many exactly timed electron pulses that arrive with a periodicity defined by an angular frequency $\omega_e$ will show a resonance at $\omega_e=\omega_0$.

In practice, our first spectroscopy protocol (Fig.~\ref{fig:preparation}a) may be realized by an electron microscope in the Lorentz mode (“in-plane” magnetic field \cite{Sugawara2019}). In addition to the differential measurement, this geometry allows for a null measurement after a $\pi/2-$pulse that eliminates any magnetic flux through the interferometer area. The required temporal resolution for this pump-probe configuration is linked to the life-time of the spin state. This condition is easily met by fast direct electron detectors which exhibit temporal resolutions of 260 ps \cite{llopart2022timepix4}.
Our second spectroscopy protocol may be realized by exploiting the strong magnetic field generated by the pole pieces of the electron microscope to align the sample's spin along the electron beam axis. Resolution of the spin precession could be achieved with an Ultrafast Transmission Electron Microscope (UTEM) which is precisely synchronized with the precession frequency. In this setup, even coherent manipulation of the spin state with timed electron interactions might become feasible \cite{Ratzel2021}.

The interferometric phase shift difference due to a differential measurement of the two spin states along the $x$-axis ($\la \osigma_x \ra = \pm 1$), 
$ 2\Delta \phi_S = 4\theta = 2 \mu_{0}e\gamma/(2 \pi d)$, depends on the strength of the magnetic dipole moment, where $\gamma=g_{s} \mu_{B}/\hbar$ is the gyromagnetic ratio (e.g. $2 \pi \cdot 28 \,\mathrm{GHz} / \mathrm{T}$ for the electron spin, with life-times in the range of 100 ns; $2 \pi \cdot 42.6 \,\mathrm{MHz} / \mathrm{T}$ for the nuclear hydrogen spin, with life times of minutes or even hours).
This leads to a phase shift difference of $\sim 1.1\times 10^{-2} \, \mathrm{mrad}$ due to a single electron spin interferometrically probed at a distance of 1 nm. Comparing to currently achievable phase sensitivities of $\sim 4.5\times 10^{-2} \, \mathrm{mrad}$, \cite{Ishikawa2023} shows that single electron spins could soon become experimentally observable. 
Rather than the elementary case of a single spin, one could study coherent ensembles of $N_S$ spins. The resulting maximal phase shift is then $\Delta \phi_S =  \pm 2N_S\theta$. Details about the derivation of this result can be found in Appendix~\ref{sec:manyspins}. 
In transmission electron microscopy, samples with a thickness of about $10 -300 \mathrm{~nm}$ are typically examined \cite{Lichte2008}, which amounts to atomic columns consisting of up to $N_S\sim 1000$ spins.  
Realistic samples will not be completely coherent and point-like. In particular, finite-temperature effects will only lead to a partial polarization of the spin sample, and the finite size will lead to additional geometric factors, as exemplified in Appendix~\ref{sec:magpolT}. In the high-temperature regime, $k_B T\gg \gamma_e\hbar B_0$, we can estimate that a fraction $\gamma\hbar B_0/2k_B T $ of the number $N_S$ of spins will constitute the sample's net magnetization probed by the electrons. For example, at a bias field of $B_0=$1.8 T, which is a standard magnetic field at the sample region in a TEM, and a temperature of 77 K, $1.6\,\%$ of electron spins will contribute to the phase signal. This goes up to $12\,\%$ at 10 K and leads to a phase shift of $1.4\,$mrad for $d=1\,$nm (see also Table \ref{tab:phaseShifts}). Measuring this phase shift in a quantum projection noise limited scheme requires only $\lesssim 2.2\cdot 10^{6}$ electrons while a typical beam current of 1.5 $\mathrm{nA}$ in scanning TEM amounts to $\sim 10^{10}$ electrons/sec. These figures are an incentive to advance electron microscopy at dilution refrigerator temperatures in the sub-Kelvin range \cite{zu2022development}.

In such cryogenic environments or by using hyperpolarisation techniques \cite{Le2023, Blanchard2021, Duckett2012} and summing over atomic “spin” columns, even the weak phase shifts due to nuclear spins should become detectable, especially since nuclear spins exhibit very long coherence times on the order of minutes to even hours \cite{de2019vivo}. A short derivation of the phase shift due to a sample of $I=1/2$ nuclear magnetic spins based on the model of classical magnetization can be found in Appendix~\ref{sec:magpolT}.

\begin{table}
    \centering
    \includegraphics[width=8cm,angle=0]{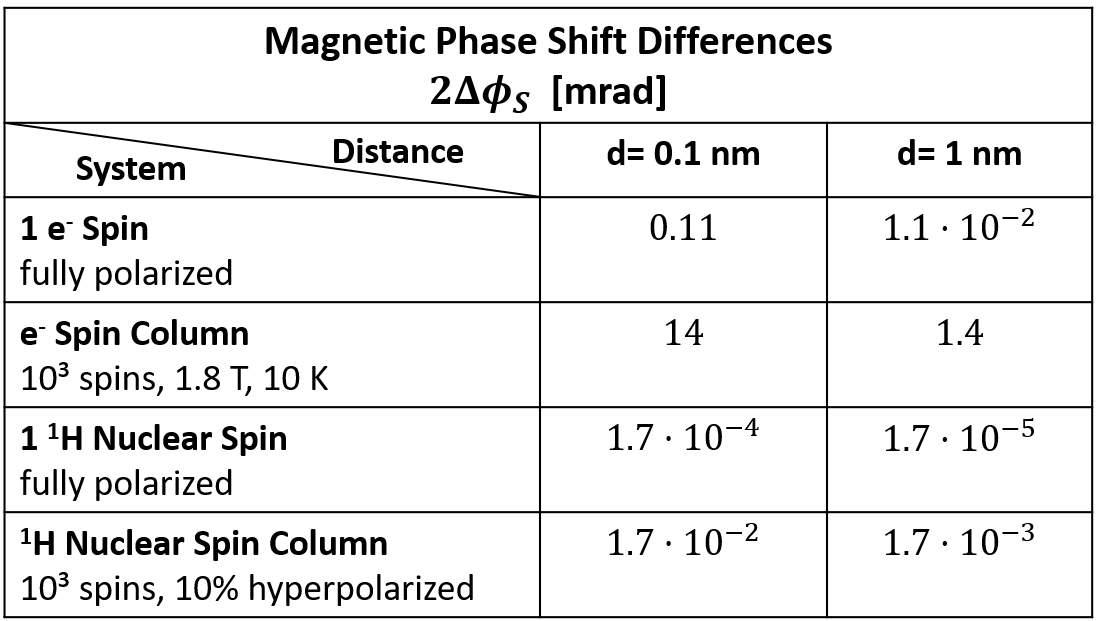}
    \caption{Estimates of the difference $2\Delta\phi_S$ of magnetic phase shifts an electron experiences due to the magnetic moment of a spin system at distance $d$ for the two spin orientations orthogonal to the interferometer plane. We compare single electron and  hydrogen nuclear spins to partially spin-polarized atomic columns of 1000 spins. For the electron spin column, we take the thermal polarization of 12\% at a magnetic bias field of 1.8 T at a temperature of 10 K. The nuclear spin column is assumed to be 10\% hyperpolarized \cite{Capozzi2015,Le2023,Blanchard2021,Duckett2012}. }
    \label{tab:phaseShifts}
\end{table}

Owing to various technical improvements in aberration correction, fast direct electron detection and highly coherent electron sources, Differential Phase Contrast (DPC) measurements reach deflection sensitivities below $25$ nrad (spatial resolution of $\sim 9\,$nm) \cite{boureau2020high}. The deflection due to a single electron spin corresponding to the gradient of the single-arm phase $\Delta\phi_S/2$ is 
$\alpha = \hbar\Delta\phi_S/(2m\gamma_L v d)$, given the $1/d$-dependence of $\Delta\phi_S$ in equation \eqref{eq:DeltaphiS} and the Lorentz factor $\gamma_L$ (see Appendix~\ref{sec:defl}). For a probe beam distance of $d=0.1\,$nm and $d=1\,$nm and electron kinetic energies of $200\,$keV, we find $\alpha \sim 110\,$nrad and $\alpha \sim 1.1\,$nrad, respectively. This implies that measurements of the deflection due to single spins may be feasible in the near future with DPC.

These promising figures encourage combining non-invasive magnetic resonance techniques with electron microscopy with wide-ranging applications from characterizing spin dynamics in battery electrodes \cite{Lu2017, Yamamoto2010} to investigating biological systems \cite{Huelga2013,Mims2021} with NMR/ESR techniques on the nanoscale.
 In principle, since the electron only needs to pass adjacent to the region of interest, radiation damage could be strongly suppressed compared to the direct examination of samples.  
Moreover, for certain optimized geometries, the electron only acquires a phase and no energy is transferred to the quantum system.
Realizing this non-invasive scheme to probe a quantum property previously inaccessible to electron microscopy techniques will provide deeper insights into the quantum nanoworld at the highest spectroscopic resolution.

\section*{Acknowledgements}

The authors thank Thomas Schachinger, Giovanni Boero, Peter Schattschneider, Antonin Jaros, Andrea Pupic, Johann Toyfl, and Santiago Steven Beltrán Romero for fruitful discussions. DR and PH acknowledge the hospitality of the Erwin Schr\"odinger Institute in the framework of their ``Research in Teams" project. DR thanks the Humboldt Foundation and the Marie Skłodowska-Curie Actions IF program for support. PH thanks the Austrian Science Fund (FWF): Y1121, P36041, P35953. This project was supported by the ESQ-Discovery Program 2019 "Quantum Klystron (QUAK)" and the FFG-project AQUTEM.

\bibliography{modulated_beam}

\onecolumngrid

\section*{Appendix}

\appendix

\section{Phase shift}
\label{sec:phase}

Since $\vec{A}$ vanishes at infinity we can express the closed path in the Aharonov-Bohm phase shift \eqref{eq:deltaPhiS_A} by two trajectories passing the dipole moment at $\pm \vec{d}=\pm(d_x\ve_x + d_y\ve_y)$ (and their closure at infinity) such that
\begin{equation}
\begin{aligned}\label{eq:deltaPhiS}
  \Delta\phi_{S} 
   & \approx  -\frac{e}{\hbar} \int_{-\infty}^\infty dz \left(A_z(-\vec{d}  + z \vec{e}_z) - A_z(\vec{d}  + z \vec{e}_z)\right)\\
  & \approx  2\frac{e}{\hbar} \int_{-\infty}^\infty dz A_z(\vec{d}  + z \vec{e}_z).
\end{aligned}
\end{equation}
where we have neglected higher order contributions of the deflection of the electron due to the magnetic moment (see next section) and used that $\vec{A}(\vec{r})=-\vec{A}(-\vec{r})$ and  $\vec{v}_0=v \vec{e}_z$. With the explicit form of the vector potential given in \eqref{eq:vecpot} and considering the case that $d_x=0$, we obtain 
\begin{equation}
\begin{aligned}
  \Delta\phi_{S} & \approx \frac{e\mu_0}{2\pi\hbar}\mu \int_{-\infty}^\infty dz \frac{d}{\left[d^2 + z^2 \right]^{3/2}} 
\end{aligned}
\end{equation}
and integration leads to equation \eqref{eq:DeltaphiS}.

\section{Deflection of electrons}
\label{sec:defl}

Electrons passing a magnetic moment are subject to the Lorentz force and thus get deflected from their free rectilinear trajectory. For the velocities and magnetic moments considered here, this deflection is tiny and needs not be accounted for in the interferometric scheme considered in the main text---provided the deflected arms still largely overlap upon recombination. Nonetheless, such tiny deflections could still be detected by a DPC measurement in the far field, as mentioned in the main text. Here, we calculate the deflection angle in the high-velocity and weak-interaction regime and show that it is approximately given by the gradient of the interferometric phase shift with respect to the distance to the sample.

The vector potential $\vec{A}$ was given in \eqref{eq:vecpot} for a magnetic dipole oriented in $\ve_x$-direction. The electron will be affected by the corresponding Lorentz force 
\begin{equation}
	\vec{F}_L = -e\vec{E} - e\vec{v}\times \vec{B} = e \partial_t\vec{A} - e\vec{v}\times(\nabla\times \vec{A}),
\end{equation}
and we can formally express the deflected electron trajectory by
\begin{equation}
    \begin{aligned}
	  \vec{p}(t) &= \vec{p}_0 + \delta\vec{p}(t) = \vec{p}_0 + \int_{-\infty}^t dt' \,\vec{F}_L(\vec{p}(t'),\vec{r}(t')) \\
	\vec{r}(t) &= \vec{r}_0{(-\infty)} + \int_{-\infty}^t dt' \frac{\vec{p}(t')}{m \sqrt{1+|\vec{p}(t')|^2/(mc)^2}}\,.
    \end{aligned}
\end{equation}
The undeflected rectilinear trajectory is given by $\vec{r}_0 (t)=\vec{d} + v_0 t\ve_z $ and $\vec{p}_0=m\gamma_Lv_0\ve_z$, where $\gamma_L=1/\sqrt{1-(v_0/c)^2}$ is the Lorentz factor. 
To leading order, the total transverse change in momentum becomes
\begin{equation}
\begin{aligned}
\label{eq:deltap_inf}
\delta \vec{p}(\infty) &\approx  \int_{-\infty}^\infty dt' \,\vec{F}_L(\vec{p}_0,\vec{r}_0(t'))	\\
 &=  e \int_{-\infty}^\infty dt' \,  v_0 \ve_z \times \left.(\nabla \times \vec{A})\right|_{\vec{r}=\vec{r}_0(t')}  \\
  &=  e\int_{-\infty}^\infty dz' \left.\left( \nabla A_z - \partial_z \vec{A}\right)\right|_{\vec{r}=\vec{d}+z'\vec{e}_z} \\ 
 &= e\int_{-\infty}^\infty dz' \left. (\vec{e}_x\partial_x + \vec{e}_y\partial_y) A_z \right|_{\vec{r}=\vec{d}+z'\vec{e}_z} \\
  & \quad - e\left[\vec{e}_x A_x(\vec{d}+z\vec{e}_z) + \vec{e}_y A_y(\vec{d}+z\vec{e}_z) \right]_{-\infty}^{+\infty} \\
  &= e\int_{-\infty}^\infty dz' \left. (\vec{e}_x\partial_x + \vec{e}_y\partial_y) A_z \right|_{\vec{r}=\vec{d}+z'\vec{e}_z}\\
  &= e \nabla_{\vec{d}} \int_{-\infty}^\infty dz  A_z(\vec{d}+z\vec{e}_z) \approx -\frac{\hbar}{2} \nabla_{\vec{d}} \Delta\phi_S
\end{aligned}
\end{equation}
Here we have used that 
$\vec{v}_0=\vec{p}_0/m\gamma_L$ and that 
$\vec{A}$ vanishes at spatial infinity.
The corresponding deflection angle is 
\begin{equation}
    \alpha = \frac{|\delta\vec{p}(\infty)|}{p_0} = \frac{\hbar }{2mv_0 \gamma_L} \sqrt{ \left(\frac{\partial \Delta \phi_S}{\partial d_x} \right)^2 + \left(\frac{\partial \Delta \phi_S}{\partial d_y} \right)^2 }.
\end{equation}

\section{Effects on position degree of freedom of electron wave function and spin state}
\label{sec:posDOFs}

When modeling the electron MZI read-out, we have assumed that the deflection induced by the magnetic moment $\vmu$ does not lead to a significant change of the electron wave packet. This is the case if the transverse width of the electron's momentum wave function is much larger than $\delta p_y$. 
If we assume that the initial wave packet is Gaussian with width $\delta y$, the final state is a shifted Gaussian with approximately the same width and their overlap is $\exp(-(\delta y\delta p_y/\hbar)^2/2)$. Hence, for a Heisenberg limited electron beam, the transverse deflection has no significant effect on the overlap of the initial and final electron wave packet if the initial wave packet is focused much more tightly than $\delta y_{\mathrm{max}}=|\partial_d\Delta\phi_S|^{-1} = 2\pi d^2/(\mu_0 e \gamma)$. For electron spin magnetic moments and electrons with kinetic energy of $200\,$keV, we find the upper limit $\delta y_{\mathrm{max}}\sim 200\,\mathrm{\mu m}$ which poses no significant limitation to experiments.

In a previous paper \cite{Ratzel2021}, some of us have studied the transitions of spin systems induced by electron pulses propagating in their vicinity which corresponds to a single arm of the MZI in Fig.\ref{fig:MZI}. We have found that, in the case that such transition occurs, the back action on the electron is negligible if $\hbar\omega_0/v \ll \hbar/(2\Delta z)$ and $d \ge 8\Delta r_\perp$, where $\Delta z$ and $\Delta r_\perp$ is the longitudinal and transverse width of the electron wave packet, respectively. The first condition essentially means that the momentum transfer to the electron corresponding to the kinetic energy exchanged with the spin system in case of transition of the latter must be much smaller than the electron's longitudinal momentum spread. The second condition ensures that flips of the electron spin appear with negligible rate. The latter condition is already covered by the assumption that the electron wave packets are well-localized on semi-classical trajectories. The first condition is fulfilled by electron wave packets in most setups.

Note that the above conditions are only significant for our theoretical model. The spectroscopic protocols provided in the main text perform measurements on eigenstates of the spin operator. Therefore, effects on the electron wave function will not lead to entanglement between electron and spin state, and our measurement protocols can be considered as back-action free.

\section{Magnetic field of the electron acting on the sample}
\label{sec:magfieldonsample}

The Aharonov-Bohm phase shift \eqref{eq:DeltaphiS} can be alternatively seen from the reverse perspective in which the moving electron induces a magnetic field that acts on the magnetic moment $\hat\vmu (t) = \mu \ovsigma (t)$ of the sample spin. The two Mach-Zehnder arm states $|L\ra$ and $|R\ra$ correspond to the electron passing by the sample at $y_L=-d$ and $y_R=d$ along the trajectory $z=v (t-t_0)$, with the time of closest proximity $t_0$. 
The respective magnetic fields at the sample location ($y,z=0$) are 
\begin{equation}
    \vB (t) = \frac{\mu_0 e \gamma v d \ve_x}{4\pi \left[ d^2 + \gamma_L^2 v^2 (t-t_0)^2 \right]^{3/2}}
\end{equation}
for $|L\ra$ and $-\vB(t)$ for $|R\ra$; see Eq.~11.152 in Ref.~\cite{jackson}. Here, $\mu_0$ denotes the vacuum magnetic permeability, $e$ the electronic charge, $v$ the electron velocity, and $\gamma_L = 1/\sqrt{1-v^2/c^2}$ the Lorentz factor. In the resulting interaction Hamiltonian 
\begin{equation}\label{eq:Hint}
    \oH_{\rm int} (t) = \hat\vmu (t) \cdot \vB (t) \left( |R\ra\la R| - |L\ra\la L| \right),
\end{equation}
we neglect the deflection of the electron trajectory caused by the interaction and assume that it cannot be resolved by a measurement on the electron. We also ignore the spin of the passing electron, which would cause a weaker, shorter-range magnetic dipole-dipole interaction (see e.g. \cite{Ratzel2021} and the discussion above).

We focus here on the quantum regime of a single spin-$1/2$ aligned with respect to a bias field $\vB_0 = B_0 \vn$ and driven by a short pulse in order to prepare a desired input state $|\psi \ra$. To this end, we replace $\vmu(t)$ by a spin operator,
\begin{equation}\label{eq:freeevo}
    \ovmu(t) = \mu\, \ovsigma(t), 
\end{equation}
that freely precesses at the Larmor frequency $\omega_0 = \mu B_0/\hbar$ about the magnetic field axis $\vn$.
An extension of our treatment to the case of collective spins is given in App.~\ref{sec:manyspins}.

We find that the interaction Hamiltonian $\oH_{\rm int} (t)$ gives a unitary state transformation that represents the passage of the electron,
\begin{align} \label{eq:U}
    \oU &= e^{-i\theta \osigma_x (t_0) } |R\ra\la R| + e^{i\theta \osigma_x (t_0) } |L\ra\la L|,   
\end{align}
The unitary describes a path-dependent phase on the electron controlled by the spin of the sample. 
Hence, the interferometric measurement will read out the $x$-projection of the sample magnetic moment through the net phase shift between the arms. The unitary also describes the measurement backaction exerted by the electron: a controlled rotation of the sample spin by an angle $\pm \theta$. This may entangle the states of sample and electron and thus result in a loss of interference visibility. Suppose the electron is split into the superposition state $|\phi\ra = (|R\ra + e^{i\phi}|L\ra)/\sqrt{2}$, with $\phi$ an adjustable external phase shift, then the combined sample-electron state becomes
\begin{equation} \label{eq:U_expanded}
    \oU |\psi \ra |\phi\ra = \cos\theta |\psi \ra |\phi\ra - i\sin\theta \osigma_x (t_0) |\psi \ra |\phi+\pi\ra.
\end{equation}
Given that $|\phi+\pi\ra$ is orthogonal to $|\phi\ra$, the combined state is entangled, unless $|\psi\ra$ is an eigenstate of $\osigma_x (t_0)$. 
To complete the measurement, the two arms of the interferometer are combined at a beam splitter.

If the sample spin is not aligned, but in an arbitrary state $|\psi\ra$, it gets rotated by the field and thus entangled with the electron path through \eqref{eq:U}. In the presence of this backaction, the simple Aharonov-Bohm description is no longer valid. Nonetheless, we can identify a net phase shift $\Delta \phi_S$ in the likelihood to detect the electron in one of the two recombined output ports,
\begin{equation}\label{eq:likelihood2}
    p_+ (\theta,\phi; t_0) = \la \psi | \la \phi | \oU^\da \left( \id \otimes |+\ra\la +| \right) \oU | \psi \ra| \phi\ra.
\end{equation}
Inserting \eqref{eq:U_expanded} leads to the simple expression \eqref{eq:likelihood}, which exhibits sinusoidal interference fringes phase-shifted by $\Delta\phi_S$ and reduced in visibility by
\begin{equation}\label{eq:visibility}
    \mathcal{V} = \sqrt{ 1 - \left[1-\la \osigma_x(t_0)\ra^2\right] \sin^2 2\theta }.
\end{equation}
At a finite temperature $T$, the likelihood \eqref{eq:likelihood2} must be averaged over the thermal distribution of spin states. For $\vn = \ve_x$ and weak interaction strengths $\theta \ll 1$, this leads to the slightly reduced visibility $\mathcal{V} \approx 1 - 8 \bar{n}_T (1-\bar{n}_T) \theta^2$, where $\bar{n}_T = 1/[1+\exp(\hbar\omega_0/k_B T)]$.

\section{Measurement Precision}
\label{sec:measprec}

Given $N_e$ electrons that are sent one by one through the Mach-Zehnder model setup of Fig.~\ref{fig:MZI} and recorded in either of the two output ports $\pm$, the achievable estimation precision of the interaction strength parameter $\theta$ is obtained from standard estimation theory as follows.   
Each single-electron measurement represents a Bernoulli trial with the probability $p_+$ in \eqref{eq:likelihood}. Its sensitivity to the parameter $\theta$ is quantified by the Fisher information (FI), 
\begin{equation}
    \mathcal{F}_{\theta}=\frac{(\partial_\theta p_+)^2}{p_+(1-p_+)}\,.
\end{equation}
Repeating the measurement with $N_e \gg 1$ electrons, the FI provides an upper bound to the achievable measurement precision, i.e., lower bound to the mean-square error $\mathrm{Var}[\theta]$ of the (unbiased) $\theta$-estimate: 
\begin{equation}\label{eq:CRB}
    \mathrm{Var}[\theta] \ge \frac{1}{N_e \mathcal{F}_{\theta}} \,.
\end{equation}
This is known as the Cram\'{e}r-Rao bound \cite{van2007parameter}. In the present case, one can understand it simply in terms of error propagation: one estimates $\theta$ from the mean count (or current) difference between the ports, $\la I\ra = N_e (2p_+ -1)$. The statistical variance of counts around the mean value is $\mathrm{Var}[I] = 4N_e p_+(1-p_+)$. Standard error propagation, valid in the limit $N_e\to\infty$ of Gaussian statistics, yields $\mathrm{Var}[\theta] \to \mathrm{Var}[I] (\partial\theta/\partial \la I \ra)^2$, which equals the right hand side of \eqref{eq:CRB}.
Inserting the explicit expression \eqref{eq:likelihood}, the bound reads as
\begin{align}\label{eq:var}
    \mathrm{Var}[\theta] &\ge   \frac{1 - \left(\cos\phi\cos2\theta - \sin\phi\sin2\theta \la \osigma_x(t_0)\ra\right)^2}{4N_e|\cos\phi\sin2\theta + \sin\phi\cos2\theta \la \osigma_x(t_0)\ra|^2}\,.
\end{align}
It has a minimum value of $1/4N_e$ at $\phi=0$, independently of $\theta$ and $\la \osigma_x(t_0)\ra$.

\section{Many spins}
\label{sec:manyspins}

We assume that $N_S$ spin systems are located at the same position in the middle of the interferometer plane. In comparison to the single-spin description of the main text, we must replace the time evolution operator by
\begin{align}
    \oU &= e^{-i\theta \sum_{m=1}^{N_S}\osigma_{x,m} (t_0) } |R\ra\la R| + e^{i\theta  \sum_{m=1}^{N_S}\osigma_{x,m} (t_0) } |L\ra\la L|
\end{align}
where $\osigma_{x,m}$ acts on the tensor product space formed by the Hilbert spaces of the individual spin systems as $\osigma_x$ on the $m$-th spin and as $\mathbb{I}$ on the others. 
Let us assume that $|\mathrm{in},N_S\ra$ is a coherent spin state, where all individual spins are in the same pure state, $|\psi,N_S\ra = |\psi\ra \otimes ... \otimes |\psi\ra $. Given the two-arm electron state $|\phi\ra$, the likelihood \eqref{eq:likelihood2} for detecting the electron in the $+$-port generalizes to
\begin{align}
      p_+ (\theta,\phi) &= \la \psi,N_S | \la \phi | \oU^\da \left( \id \otimes |+\ra\la +| \right) \oU | \psi,N_S \ra| \phi\ra = \frac{1 + \mathrm{Re}(e^{i\phi}D_S)}{2}, \nonumber \\
      \text{with } \,\, D_S &= \la \psi,N_S |e^{2i\theta \sum_{m=1}^{N_S}\osigma_{x,m} (t_0) } | \psi,N_S \ra \, .
\end{align}
In analogy to the specific protocols described in the main text, we consider first the case 
 in which $|\psi\ra$ is the eigenstate of $\sigma_x$ with eigenvalue $1$ such that
\begin{align}
      D_S &= e^{i2N_S\theta}\,.
\end{align}
This implies a phase shift $\Delta\phi_S= 2N_S\theta$, a visibility $\mathcal{V}=1$, and the Cram\'{e}r-Rao precision bound $\mathrm{Var}[\theta] \ge 1/(4N_e N_S^2)$.

For the second case, we assume that 
$\la \psi|\osigma_{x}(t_0)|\psi\ra$ vanishes. 
Since $\la \psi|(\osigma_{x}(t_0))^{2n+1} |\psi\ra = 0$ and $\la \psi|(\osigma_{x}(t_0))^{2n} |\psi\ra = 1$, we obtain 
\begin{align}
      D_S 
      &= \la \psi,N_S| \sum_n \frac{(i\theta)^n}{n!} \left(\sum_{m=1}^{N_S} \osigma_{x,m} (t_0) \right)^n|\psi,N_S\ra \nonumber\\
      &= \la \psi,N_S|  \sum_n \frac{(i\theta)^n}{n!} \sum_{k_1+...+k_{N_S}=n}\frac{n!}{k_1!\cdot\cdot\cdot k_{N_S}!} (\osigma_{x,1} (t_0))^{k_1}\cdot\cdot\cdot(\osigma_{x,N_S} (t_0))^{k_{N_S}}  |\psi,N_S\ra   \nonumber\\
      &= \sum_n \frac{(-1)^n\theta^{2n}}{(2n)!} \sum_{k_1+...+k_{N_S}=n}\frac{(2n)!}{(2k_1)!\cdot\cdot\cdot (2k_{N_S})!} = \sum_n \sum_{k_1+...+k_{N_S}=n}\frac{(-1)^{k_1+...+k_{N_S}}\theta^{2k_1+...+2k_{N_S}}}{(2k_1)!\cdot\cdot\cdot (2k_{N_S})!} \nonumber\\
      &= \sum_{k_1}\frac{(-1)^{k_1}\theta^{2k_1}}{(2k_1)!}\cdot\cdot\cdot \sum_{k_{N_S}}\frac{(-1)^{k_{N_S}}\theta^{2k_{N_S}}}{(2k_{N_S})!} = \cos^{N_S}(\theta)
\end{align}
This implies a vanishing phase shift and visibility of $\mathcal{V}=\cos^{N_S}(2\theta)$, and we find
\begin{align}
    \mathrm{Var}[\theta] \ge  \frac{1-\cos^{2N_S}(2\theta)}{N_e(2N_S\sin(2\theta)\cos^{N_S-1}(2\theta))^2}
\end{align} 
which for small $\theta$ becomes $\mathrm{Var}[\theta] \ge 1/(4N_e N_S)$.

In summary, we find that the induced phase on the electron interferometer adds coherently and the precision of inference of the interaction strength increases with $N_S$, while it increases with $\sqrt{N_S}$ for the decoherence process. In the limit of large $N_S$, the phase becomes the only relevant observable and the result can be compared to the situation of a classical (precessing) magnetic dipole moment inducing a phase in the electron interferometer.

\section{Phase shift due to magnetic polarization at finite temperature}
\label{sec:magpolT}

Until now, we have restricted our considerations to spin systems that are point-like in comparison to the extension of the electron interferometer. To understand what changes if multiple spin systems are distributed on length scales of a similar size as the interferometer, here we restrict to a classical model of the spin systems. As we found above, this should provide accurate results if the interaction with the electron is stroboscopic, that is the duration of significant interaction is much shorter than the spin precession period, and a sufficient number of spins are orthogonal to the electron interferometer plane at the times of interaction with the electron pulses such that the phase shift dominates over decoherence.

Then, we model the spin system as a paramagnetic sphere with homogeneous, isotropic magnetic susceptibility $\chi$ that is placed in a homogeneous magnetic field $\mathbf{B}_0$. The resulting homogeneous polarization is (see \cite{jackson} Section 5.11)
\begin{equation}
    \mathbf{M}=\frac{3}{\mu_0}\left(\frac{\mu-\mu_0}{\mu+2\mu_0}\right) \mathbf{B}_0 \approx \chi \mathbf{B}_0/\mu_0\,
\end{equation}
where $\mu=(1+\chi)\mu_0$ and we assumed $\chi\ll 1$ for the approximate result. The magnetic field induced by the magnetization inside the sphere is  
\begin{equation}
    \mathbf{B}'_{in}= \frac{2\mu_0}{3} \mathbf{M} = \frac{2}{3}\left(\frac{\mu-\mu_0}{\mu+2\mu_0}\right) \mathbf{B}_0 \approx \frac{2}{3}\chi\mathbf{B}_0\,.
\end{equation}
If we assume that the spin vectors corresponding to this magnetization are subjected to an ideal $\pi/2$-pulse, the resulting precession leads to the oscillating magnetic field with amplitude $B'_{in}=|\mathbf{B}'_{in}|=2\chi B_0/3$. Placing the sphere in the interferometer such that the cross section of the sphere in the interferometer plane is maximized, the maximal magnetic flux that can be obtained  (if the electron interferometer encloses the whole sphere and just the sphere) is $\Phi=\pi R^2 B'_{in}$, where $R$ is the radius of the sphere.

In a bias field $B_0$ the induced energy split between the spin states of an electron is $\gamma_e\hbar B_0$, where $\gamma_e$ is the electron gyromagnetic ratio. In the following, we consider a paramagnetic electron-spin medium with a temperature such that $k_B T\gg \gamma_e\hbar B_0$. In that case, we find for the magnetic susceptibility (with spin $I=1/2$ and $\gamma=\gamma_e$) \cite{boero2000integrated}
\begin{equation}
    \chi = \mu_0\frac{n_S \gamma_e^2\hbar^2 I(I+1)}{3k_B T} =  \mu_0\frac{n_S \gamma_e^2\hbar^2}{4k_B T}\,,
\end{equation}
where $n_S=N_S/V$ is the electron spin density, $N_S$ is the number of electron spins in the sphere and $V$ is the volume of the sphere. We find for the magnetic flux
\begin{equation}
    \Phi \approx  N_S \frac{g_S \mu_0 \mu_B}{R} \frac{\gamma_e\hbar B_0}{8k_B T} \,.
\end{equation}
For the magnetic phase in the electron MZI, we find
\begin{equation}
    \Delta\phi_S = -\frac{e}{\hbar} \Phi \approx - N_S g_S \frac{r_e}{R} \frac{\pi}{2} \frac{ \gamma_e\hbar B_0}{2 k_B T}\,,
\end{equation}
with $r_e$ the classical electron radius and $\gamma_e=2 \pi \cdot 28$ GHz the electron gyromagnetic ratio.
Besides the Boltzmann factor and the numerical factor $\pi/2$, we recover the results of the previous section. 

Of course, similar calculations as for the interaction of a free electron with an electron spin can be made for the spin of a nucleus. We simply consider $ \mu = \mu_N g_I/2 $, where $\mu_N = e\hbar/(2m_p)$ is the nuclear magneton and $g_I$ is the total nuclear spin g-factor. For simplicity, we assume that the nuclear spin is also $1/2$ and we use the representation in terms of the Pauli matrices as in the previous case writing $\hat{\vec{\mu}}=-\mu_N g_I\vec{\sigma}/2$. Accordingly, we recover all above equations with the replacement of the coupling parameter $\theta$ by
\begin{equation}
    \theta_I= \frac{\mu_0 e \mu_N g_I}{4\pi\hbar d} = \frac{\mu_0 e}{d}\frac{\gamma_N}{4\pi}  \,,
\end{equation}
where $\gamma_N$ is the nuclear gyromagnetic ratio which depends on the atoms that are probed. The resulting phase shift is
\begin{equation}
    \Delta\phi_S = 2\theta_I = \frac{\mu_0 e}{d}\frac{\gamma_N}{2\pi}  \,.
\end{equation}
In analogy to the above analysis, we can also calculate the phase shift due to a classical magnetization corresponding to nuclear spins for $k_B T\gg \hbar\gamma_N B_0$ \cite{boero2000integrated}
\begin{equation}
    \chi = \mu_0\frac{n_S \gamma_N^2\hbar^2 I(I+1)}{3k_B T}   \,,
\end{equation}
and the  maximal phase due to a magnetized sphere (if the electron interferometer encloses the whole sphere and just the sphere) becomes
\begin{equation}
    \Delta\phi_S = -\frac{e}{\hbar} \Phi \approx  - N_S \frac{e \mu_0 }{R}\frac{\gamma_N}{2\pi} \frac{\pi}{3}\frac{\hbar\gamma_N B_0}{k_B T} I(I+1)   \,,
\end{equation}
and for $I=1/2$, we find
\begin{equation}
    \Delta\phi_S = - N_S \frac{e \mu_0 }{R}\frac{\gamma_N}{2\pi} \frac{\pi}{2}\frac{\hbar\gamma_N B_0}{2k_B T} \,
\end{equation}
recovering $2N_S\theta_I $ up to the Boltzmann factor and the numerical factor $\pi/2$.

\end{document}